# VARIABILITY OF LOCATION MANAGEMENT COSTS WITH DIFFERENT MOBILITIES AND TIMER PERIODS TO UPDATE LOCATIONS

E. Martin and R. Bajcsy

Department of Electrical Engineering and Computer Science
University of California, Berkeley
California, USA
`emartin@eecs.berkeley.edu`

## ABSTRACT

*In this article, we examine the Location Management costs in mobile communication networks utilizing the timer-based method. From the study of the probabilities that a mobile terminal changes a number of Location Areas between two calls, we identify a threshold value of 0.7 for the Call-to-Mobility Ratio (CMR) below which the application of the timer-based method is most appropriate. We characterize the valley appearing in the evolution of the costs with the timeout period, showing that the time interval required to reach 90% of the stabilized costs grows with the mobility index, the paging cost per Location Area and the movement dimension, in opposition to the behavior presented by the time interval that achieves the minimum of the costs. The results obtained for CMRs below the suggested 0.7 threshold show that the valley appearing in the costs tends to disappear for CMRs within [0.001, 0.7] in one-dimensional movements and within [0.2, 0.7] in two-dimensional ones, and when the normalized paging cost per Location Area is below 0.3.*

## 1. INTRODUCTION

Within Mobility Management for mobile communications networks, Location Management enables the roaming of the user in the networks, and the two main procedures involved in this task are: location update and paging. Location update provides the network with information about the location of the user (in terms of Location Areas), while paging is responsible for delivering incoming calls to the user. In terms of signaling costs, these two procedures present opposite behaviors, and the optimization of the trade-off in the overall Location Management costs has become a key research topic [1-5]. Several algorithms have been proposed in the literature to optimize these signaling costs, with the three main ones being: i) movement-based [6], ii) distance-based [7] and iii) timer-based [8], according to the main parameter used to control the location update procedure: a number of movements (usually cell-crossings), a distance from the origin of the movement, or a certain time threshold, respectively. The performance of the movement-based and distance-based algorithms for Location Management has received special attention by researchers, but exact numerical studies have not been carried out for the timer-based algorithm [9]. Consequently, there is a need for an exhaustive numerical analysis of the timer-based method.
In this sense, reference [10] presents a location update scheme employing a timer, tackling its analytical optimization by modeling the timer as a function of environmental conditions (traffic generation, mobility and cost of location update). Within the context of Mobile IP, researchers have also shown the need for a detailed analysis of the influence of residence time on the different steady state probabilities [11]. Reference [12] evaluates Cellular IP mobility tracking procedures, showing the importance of the appropriate selection of the timer values. In fact, Mobility Management in Cellular IP is based on the two states that the mobile host can take: idle or active; and for the idle state, the Location Management strategy used is a combination between the classical scheme (triggering updates when the borders of predetermined sets of cells are crossed) and the timer-based method.

Regarding multi-access networks, reference [13] presents an analytical study of Mobility Management focusing on the importance of the dwell-timer parameter, whose optimal value is shown to depend on the particular scenario and parameters accounting for the user's mobility (e.g. velocity). In fact, reference [14] shows that the knowledge of the speed of the user can play a critical role to successfully improve the performance of location-based services. In this sense, recent research shows the feasibility of obtaining precise kinematic information about the users processing their acceleration with the wavelet transform [15-17]. Actually, the wavelet transform can be successfully applied to analyze the waist acceleration obtained from the sensors embedded in the smart phones and obtain the velocity of the user [18], and this is possible for a wide range of time windows as shown in reference [19]. In the same sense, other sensors (e.g., magnetometer) embedded in smart phones can also help enhance the accuracy in the localization solution [20]. Additionally, the radios in the smart phones can be leveraged to improve the localization solution by means of fingerprinting techniques [21-22].

The employment of a timer value and a movement list in order to achieve reductions in the Location Management costs has also been recently proposed for a dynamic anchor-area scheme [23]. As another example, reference [24] focuses on the timer-based algorithm for Location Management, assigning dynamically individualized Location Areas to each mobile user, based on their mobility rates and a timer. In GPRS, a time interval is defined to determine when to switch from cell to Routing Area while tracking a mobile terminal. In this sense, recent studies on Mobility Management for GPRS have highlighted the problem of the possible loss of synchronization between the mobile terminal and the serving GPRS support node, proposing analytic and simulation models to investigate different alternatives to mitigate the referred issue [25-26]. Reference [8] proposes a new timer-based algorithm using two timer thresholds in order to tackle the burstiness of the data in packet-switched networks, showing that it can outperform conventional Location Management strategies in terms of signaling costs. Within the packet-switched domain of UMTS, a three-level Location Management strategy is used [27]: while the mobile station is not engaged in a communication, it is tracked down at the Routing Area level; when a communication is established, the mobile is tracked down at cell level while packets are being transmitted, and it is switched to UTRAN Registration Area (URA) level during the idle periods of the communication. The objective of this three-level strategy is the minimization of the Location Management signaling costs and power consumption, and one of the possible approaches to trigger the transitions from cell level to the URA and Routing Area levels is the use of timers.

Other examples in this field include reference [28], which proposes a location caching scheme, showing the importance of an appropriate selection of a timer value for the reduction of signaling costs. In fact, the cache timer value represents the most important control parameter in their scheme. Subsequently, the authors provide a closed-form expression for the optimum timer value based on user mobility.

In this paper we examine the Location Management costs provided by the timer-based method, with special emphasis on the characterization of the valley appearing at early stages, through measurements of several parameters. The analysis is carried out for one and two-dimensional movements. The rest of this paper is organized a follows: Section 2 presents a discussion on the Call-to-Mobility Ratio (CMR) values below which the application of the timer-based method is most appropriate. Section 3 is devoted to the characterization of the costs for one-dimensional movements while Section 4 reviews the two-dimensional case and Section 5 compares the results. The paper is concluded in Section 6.

## 2. DISCUSSION ON THE MOST APPROPRIATE CONDITIONS FOR THE APPLICATION OF THE TIMER-BASED METHOD

Considering a general distribution for the residence time of a mobile terminal in a Location Area, with mean $tr$, the probability of crossing a number of Location Areas between consecutive calls can be expressed in terms of that distribution [29]. This relationship is achieved through the Laplace-Stieltjes Transform, which can be calculated for different types of distributions [30]. In particular, for the exponential residence time distribution, commonly used in existing research [31-32], the probability of crossing $N$ areas between consecutive calls is expressed as:

$$Pcross(N) = \frac{(1/tr)^N \lambda_p}{[(1/tr) + \lambda_p]^{N+1}}, \qquad (1)$$

where $\lambda_p$ represents the call arrival rate. For the constant distribution, the referred probabilities are given by [30]:

$$Pcross(N) = \frac{1}{tr\lambda_p}\left[1 - e^{-tr\lambda_p}\right]^2 e^{-tr\lambda_p(N-1)} \qquad (2)$$

$$Pcross(0) = 1 - \frac{1}{tr\lambda_p}\left(1 - e^{-tr\lambda_p}\right). \qquad (3)$$

And for the uniform distribution [30]:

$$Pcross(N) = \frac{1}{tr\lambda_p}\left[1 - \frac{1}{2tr\lambda_p}\left(1 - e^{-\frac{2}{tr\lambda_p}}\right)\right]^2 \times \left[\frac{1}{2tr\lambda_p}\left(1 - e^{-\frac{2}{tr\lambda_p}}\right)\right]^{N-1} \qquad (4)$$

$$Pcross(0) = 1 - \frac{1}{tr\lambda_p}\left[1 - \frac{1}{2tr\lambda_p}\left(1 - e^{-2tr\lambda_p}\right)\right] \qquad (5)$$

Next, the evolution of these probabilities will be studied for different CMRs. This evolution for two particular values of the CMR is presented in Figures 1 and 2.

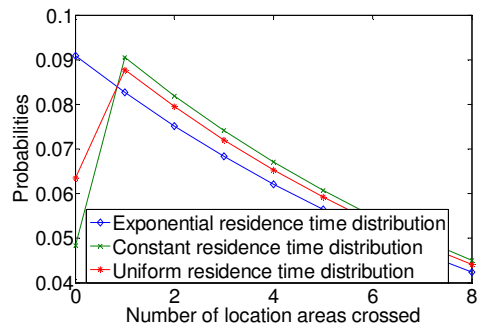

Figure 1. Probabilities of changing Location Areas between two calls for CMR=0.1.

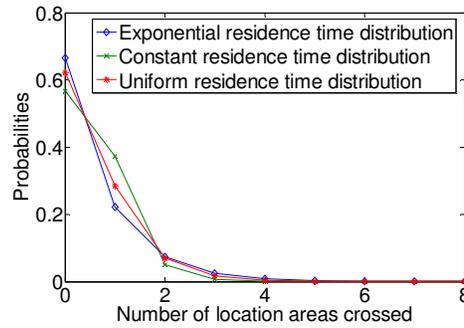

Figure 2. Probabilities of changing Location Areas between two calls for CMR=2.

For CMRs above 5, the probability that a mobile terminal does not change Location Area between two calls, $Pcross(0)$, approximately meets at the same point for the three analyzed distributions. For CMRs below 5, there is a trend for the constant distribution to have a value of $Pcross(0)$ lower than the uniform distribution, and this one, in its turn, lower than the exponential distribution. $Pcross(0)$ for the exponential distribution keeps the most predictable behavior in conjunction with the rest of samples, while in the case of the uniform and constant distributions, $Pcross(0)$ seems to lose connection from the rest of the samples for CMRs lower than 6, with more emphasis for the constant distribution than for the uniform distribution. This particularity keeps an approximately constant proportion for CMRs lower than 0.1 ($Pcross(0)$ is approximately 65% of $Pcross(1)$ for the uniform distribution, and 50% of $Pcross(1)$ for the constant distribution).

For CMR figures above 50, $Pcross(0)$ presents a much higher value than the rest of the probabilities for the three analyzed distributions. In particular, $Pcross(0)$ for these CMRs tends to stabilize at 1, while the rest of the probabilities tend to decrease their values, with the $Pcross(0)/Pcross(1)$ quotient approaching the value of the CMR. The abruptness of this leap between $Pcross(0)$ and $Pcross(1)$ increases with the CMR.

As the CMR drops to around 20, the probability to change one Location Area increases to approximately 0.05, but the probabilities to change two or more Location Areas still remain negligible. For CMRs below 10, the probability to change one Location Area keeps rising to around 0.3 for a ratio of 1, and the probabilities to cross two or even more Location Areas keep on growing, approaching 0.1. When the CMR falls below 1, the probabilities to cross more than one Location Areas rise considerably. In this sense, for CMRs below 0.5, $Pcross(0)$ decreases importantly, and its leap with $Pcross(1)$ and the rest of probabilities diminishes, at the expense of bringing down the average probability values. In particular, these average probability values tend to decrease at the same rate of the CMR, and the different consecutive probabilities tend to acquire closer values, so that the probabilities of crossing $N$ or $N+1$ Location Areas are approximately identical. In these conditions, the difference between $Pcross(0)$ and $Pcross(N)$ tends to diminish even for large values of $N$, and the lower the CMR, the larger $N$ can be.

From the previous results, it can be concluded that for CMRs above unity, $Pcross(0)$ tends to be much higher than $Pcross(1)$ or $Pcross(N)$ in general, in such a way that the amount of updates originated from arriving calls would make the application of the timer-based method lose efficiency. Thus, considering lower CMRs, a threshold can be established around 0.7, for which value, $Pcross(0)$ approximately equals $Pcross(1)$ regardless of the mobile terminal's residence time distribution. For CMRs below 0.7, the absolute difference between $Pcross(0)$, $Pcross(1)$ and $Pcross(N)$ in general is low enough as to consider feasible the application of the timer-based algorithm.

Consequently, in this paper focus will be put on low CMRs, namely ranging from $10^{-5}$ to 0.7. In particular, this interval will be split into two (approximated through the variation of the mobility index $r$ in [150, 50000] and [1.5, 14]), for a better understanding of the behavior of the costs with different values of the parameters.

## 3. STUDY OF ONE-DIMENSIONAL MOVEMENTS

Considering a Poisson process for the incoming calls with rate $\lambda_p$ and taking for simplicity purposes the cost of each location update as unity and the paging cost per Location Area as $P$, the expression for the mean Location Management costs for a one-dimensional movement is given by [33]:

$$\eta(t) = \frac{\lambda_p}{1-e^{-t}} \left( r^{1/2} P \times [erf(\sqrt{t}) - 2\sqrt{\frac{t}{\pi}} e^{-t}] + e^{-t} \right) \quad (6)$$

In (6), $r$ represents the mobility index defined as the quotient between the diffusion constant, $Dif$, and $\lambda_p$:

$$r = \frac{Dif}{\lambda_p} \quad (7)$$

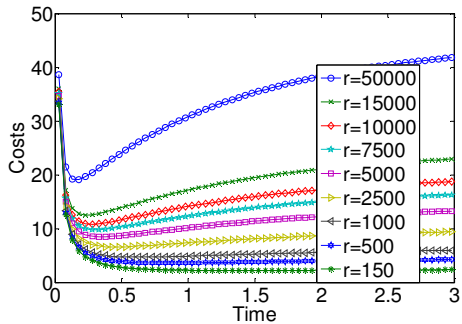

Figure 3. Location Management costs for timer-based method. $P$=0.2, $r \in$ [150, 50000]

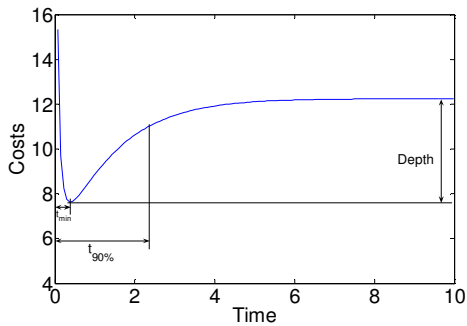

Figure 4. Characterization of the valley of costs provided by the timer-based method.

Considering the evolution of the costs for values of $r$ ranging from 150 to 50000, which correspond to CMRs between $6.7 \times 10^{-3}$ and $2 \times 10^{-5}$ respectively, it can be noticed that before a stable value is achieved, a valley appears at the beginning. In order to characterize this valley, measurements should be taken of its depth, the percentage of this depth with respect to the final stable value, the time length to reach the minimum, $t_{min}$, and the time length to reach 90% of the stable value, represented as $t_{90\%}$. As $t_{min}$ has already been studied in the past [34], we will concentrate on the other parameters.

The stable value that the costs will take, *SV*, can be easily obtained as the limit of the costs as time approaches $\infty$ :

$$SV = \lim_{t \to \infty} \frac{\lambda_p}{1-e^{-t}} \left( r^{1/2} P \times [erf(\sqrt{t}) - 2\sqrt{\frac{t}{\pi}} e^{-t}] + e^{-t} \right) = r^{1/2} \lambda_p P \qquad (8)$$

The percentage of the minimum, $P_m$, will be calculated as:

$$P_m = \frac{Depth}{SV} \cdot 100 (\%) \qquad (9)$$

Table 1. Measurements of valleys for P=0.1 for higher values of *r*.

| r | SV | Minimum | Depth | $P_m$ | $t_{90\%}$ |
|---|---|---|---|---|---|
| 150 | 1.224 | 1.194 | 0.03 | 2.45098 | - |
| 500 | 2.236 | 2.038 | 0.198 | 8.855098 | - |
| 1000 | 3.16 | 2.718 | 0.442 | 13.98734 | 1.98 |
| 2500 | 5 | 3.906 | 1.094 | 21.88 | 2.195 |
| 5000 | 7.071 | 5.087 | 1.984 | 28.05827 | 2.288 |
| 7500 | 8.66 | 5.932 | 2.728 | 31.50115 | 2.327 |
| 10000 | 10 | 6.577 | 3.423 | 34.23 | 2.3485 |
| 15000 | 12.24 | 7.625 | 4.615 | 37.70425 | 2.35 |
| 50000 | 22.36 | 11.79 | 10.57 | 47.27191 | 2.37 |

Table 2. Measurements of valleys for P=0.2 for higher values of *r*.

| r | SV | Minimum | Depth | $P_m$ | $t_{90\%}$ |
|---|---|---|---|---|---|
| 150 | 2.448 | 2.201 | 0.247 | 10.08987 | 1.45 |
| 500 | 4.472 | 3.587 | 0.885 | 19.7898 | 2.16 |
| 1000 | 6.32 | 4.679 | 1.641 | 25.96519 | 2.265 |
| 2500 | 10 | 6.577 | 3.423 | 34.23 | 2.3484 |
| 5000 | 14.142 | 8.509 | 5.633 | 39.83171 | 2.3862 |
| 7500 | 17.32 | 9.779 | 7.541 | 43.53926 | 2.402 |
| 10000 | 20 | 10.85 | 9.15 | 45.75 | 2.4126 |
| 15000 | 24.48 | 12.65 | 11.83 | 48.32516 | 2.4161 |
| 50000 | 44.72 | 19.01 | 25.71 | 57.49106 | 2.4563 |

Table 3. Measurements of valleys for P=0.3 for higher values of *r*.

| r | SV | Minimum | Depth | $P_m$ | $t_{90\%}$ |
|---|---|---|---|---|---|
| 150 | 3.672 | 3.069 | 0.603 | 16.42157 | 2.0392 |
| 500 | 6.708 | 4.889 | 1.819 | 27.11688 | 2.2755 |
| 1000 | 9.48 | 6.33 | 3.15 | 33.22785 | 2.3307 |
| 2500 | 15 | 8.852 | 6.148 | 40.98667 | 2.3913 |
| 5000 | 21.213 | 11.33 | 9.883 | 46.58936 | 2.4161 |
| 7500 | 25.98 | 13.24 | 12.74 | 49.03772 | 2.427 |
| 10000 | 30 | 14.79 | 15.21 | 50.7 | 2.4338 |
| 15000 | 36.72 | 16.72 | 20 | 54.46623 | 2.4344 |
| 50000 | 67.08 | 25.43 | 41.65 | 62.09004 | 2.456 |

Table 4. Measurements of valleys for P=0.1 for higher values of $r$.

| r | SV | Minimum | Depth | $P_m$ | $t_{90\%}$ |
|---|---|---|---|---|---|
| 150 | 11.016 | 7.07 | 3.946 | 35.82062 | 2.3533 |
| 500 | 20.124 | 10.9 | 9.224 | 45.83582 | 2.402 |
| 1000 | 28.44 | 14.23 | 14.21 | 49.96484 | 2.4227 |
| 2500 | 45 | 19.09 | 25.91 | 57.57778 | 2.4485 |
| 5000 | 63.639 | 24.44 | 39.199 | 61.59588 | 2.4553 |
| 7500 | 77.94 | 28.55 | 49.39 | 63.36926 | 2.4585 |
| 10000 | 90 | 32.01 | 57.99 | 64.43333 | 2.4608 |
| 15000 | 110.16 | 37.81 | 72.35 | 65.6772 | 2.4563 |
| 50000 | 201.24 | 63.93 | 137.31 | 68.23196 | 2.4708 |

Measurements show that for a fixed value of $r$, the larger $P$, the deeper the valley becomes. And for the same values of $P$, the greater the $r$ value, again the deeper the valley. In this sense, for $P=0.1$, $P_m$ ranges between 2.45% and nearly 50% when $r$ rises from 150 to 50000, and these percentages increase with $P$, till the point of ranging from 35% to nearly 70% for $P=0.9$, for the same rises in $r$.

The time interval to reach 90% of the final stable value varies slightly with $r$ and $P$, and the larger these parameters, the longer that time interval. However, most of the values for $t_{90\%}$ concentrate around 2.4 time units for the sets of values of $r$ and $P$ referred.

For CMRs ranging between 0.073 and 0.7, which imply $r$ values from approximately 14 down to 1.5 respectively, the valley's depth decreases, and the costs are remarkably lower than for the previous CMRs, considering a fixed value of $P$. Within these CMRs, for values of $P$ below 0.3, $P_m$ is always under 2% regardless of $r$, and when the value of $P$ is low enough, approximately 0.1, there is no valley.

Table 5. Measurements of valleys for P=0.3 for lower values of $r$.

| r | SV | Minimum | Depth | $P_m$ | $t_{90\%}$ |
|---|---|---|---|---|---|
| 1.5 | 0.367423 | 0.3674 | 2.34614E-05 | 0.006385 | - |
| 2.5 | 0.474342 | 0.4743 | 4.1649E-05 | 0.00878 | - |
| 3.5 | 0.561249 | 0.561 | 0.000248608 | 0.044296 | - |
| 4.5 | 0.636396 | 0.6356 | 0.000796103 | 0.125096 | - |
| 5.5 | 0.703562 | 0.7019 | 0.001662364 | 0.236278 | - |
| 6.5 | 0.764853 | 0.7619 | 0.002952927 | 0.386078 | - |
| 7.5 | 0.821584 | 0.817 | 0.004583836 | 0.557927 | - |
| 8.5 | 0.874643 | 0.8681 | 0.006542784 | 0.748052 | - |
| 9.5 | 0.924662 | 0.916 | 0.0086621 | 0.936785 | - |
| 10.5 | 0.972111 | 0.961 | 0.011111105 | 1.142987 | - |
| 11.5 | 1.017349 | 1.004 | 0.013349497 | 1.312184 | - |
| 12.5 | 1.06066 | 1.044 | 0.016660172 | 1.570736 | - |
| 13.5 | 1.10227 | 1.083 | 0.019270384 | 1.748245 | - |

Table 6. Measurements of valleys for P=0.9 for lower values of *r*.

| r | SV | Minimum | Depth | $P_m$ | $t_{90\%}$ |
|---|---|---|---|---|---|
| 1.5 | 1.10227 | 1.083 | 0.01927 | 1.748245 | - |
| 2.5 | 1.423025 | 1.371 | 0.052025 | 3.655941 | - |
| 3.5 | 1.683746 | 1.594 | 0.089746 | 5.330129 | - |
| 4.5 | 1.909188 | 1.779 | 0.130188 | 6.81904 | - |
| 5.5 | 2.110687 | 1.94 | 0.170687 | 8.086802 | - |
| 6.5 | 2.294559 | 2.084 | 0.210559 | 9.176439 | - |
| 7.5 | 2.464752 | 2.213 | 0.251752 | 10.21407 | - |
| 8.5 | 2.623928 | 2.332 | 0.291928 | 11.12562 | 1.6956 |
| 9.5 | 2.773986 | 2.441 | 0.332986 | 12.00389 | 1.7911 |
| 10.5 | 2.916333 | 2.544 | 0.372333 | 12.76717 | 1.8567 |
| 11.5 | 3.052048 | 2.642 | 0.410048 | 13.43519 | 1.9054 |
| 12.5 | 3.181981 | 2.732 | 0.449981 | 14.14152 | 1.9439 |
| 13.5 | 3.306811 | 2.818 | 0.488811 | 14.78195 | 1.9766 |

The time intervals required to reach 98% of the final stable value, $t_{98\%}$, are an average of 10% shorter for the subset of higher CMRs than for the subset of lower ones. Although the length of these intervals varies with *P*, the typical values range between 3.8 and 4.5 time units. Regarding $t_{90\%}$ for CMRs $\in$ [0.073, 0.7], this parameter is only considerable for large enough values of *P*. Within these cases, for example for *P*=0.9, it only makes sense to measure $t_{90\%}$ for values of *r* larger than 8.5. Measurements show that the length of the $t_{90\%}$ interval increases with *r*. In average, this $t_{90\%}$ is slightly lower than 2 time units for CMRs $\in$ [0.073, 0.7], decreasing by approximately 15% when *r* is reduced from 13.5 to 8.5.

Regarding the time intervals required to reach the minimum of the costs when the valleys appear, their length decreases with rises in *r* and *P*.

## 4. STUDY OF TWO-DIMENSIONAL MOVEMENTS

Despite the difficulty to derive analytical expressions for the timer-based method considering two-dimensional movements, the mean cost for this case can be approximated by [33]:

$$\eta(t) = \lambda_p rP\pi + \frac{\lambda_p e^{-t}}{1-e^{-t}}(1 - r\pi Pt) \tag{10}$$

Again the stable value that the costs will take is easily obtainable calculating the limit of the costs as time approaches $\infty$:

$$SV = \lim_{t \to \infty}\left(\lambda_p rP\pi + \frac{\lambda_p e^{-t}}{1-e^{-t}}(1 - r\pi Pt)\right) = \lambda_p rP\pi \tag{11}$$

In general, the behavior of the costs for the two-dimensional movements follows similar patterns as for the one-dimensional case. Regarding the time intervals to reach the minimum of the costs, those timeout values for the two-dimensional case also diminish as *r* and *P* increase.

## 5. COMPARISON OF THE COSTS ACCORDING TO THE MOVEMENT DIMENSIONS: RESULTS

Focusing on the movement dimension, values of $r \in [0.1014, \infty)$ and $P \in [0.1, 0.9]$ always bring costs for the two-dimensional processes larger than for the one-dimensional ones. However, that difference in the costs declines with falls in $r$, for a fixed value of $P$. In fact, as $r$ drops below 0.1013, the stable values for the one-dimensional movements go above the two-dimensional ones.

Considering values of $P$ under 0.3, the valley in the costs becomes perceptible only for values of $r$ approximately above 1000 in one-dimensional movements, or values of $r$ over 5 in two-dimensional processes. For fixed values of $r$ and $P$, the timeout period $t_{min}$ becomes longer for one-dimensional movements than for two-dimensional ones.

Figures 5 and 6 present the evolution of the costs for $r$ values of 14, 1400 and 14000 for one and two dimensions of the movement, showing the increase of the costs for the two-dimensional case in comparison with the one-dimensional one, for which case the costs can not be observed in Figure 5 because they are stuck to the abscissa axis.

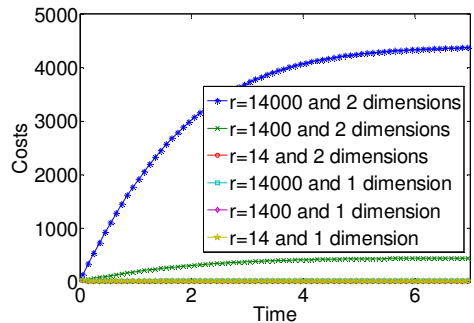

Figure 5. Comparison between 1 and 2 dimensions for large values of $r$ and $P$=0.1. Global plot.

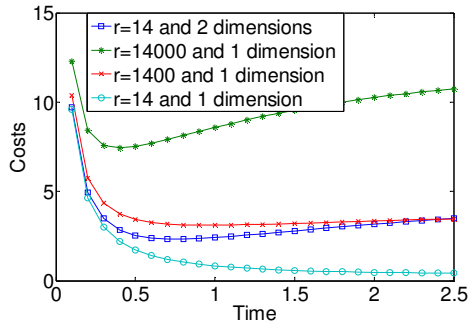

Figure 6. Comparison between 1 and 2 dimensions for large values of $r$ and $P$=0.1. Zoom of origin.

In Figure 6 a clearer comparison is shown for $r$=14, and it can be observed that the difference between the one-dimensional and two-dimensional cases is not as big as with large values of $r$. In the same sense, Figure 7 illustrates a comparison of the costs with the dimension of the movement for values of $r$ ranging between 1.4 and 8.

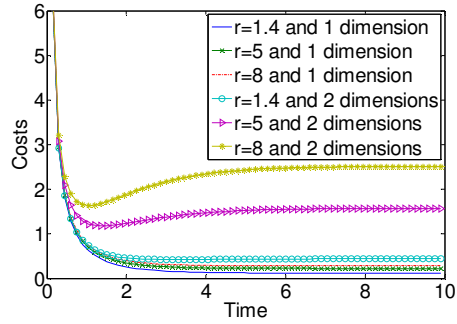

Figure 7. Comparison between 1 and 2 dimensions for low values of *r* and *P*=0.1

Again, the costs associated to the two-dimensional movement are larger than those of the one-dimensional movement, although that difference diminishes as the value of *r* decreases. In Figure 7, it can be noticed that for the selected values of *r* and *P*, the valley appears only for the two-dimensional movements. As happened with the one-dimensional case, for the two-dimensional one, the larger the values of *r* and *P*, the more important the valley becomes.

For the values of *r* analyzed, and for the cases in which the $t_{90\%}$ parameter is effective, this parameter in the two-dimensional version is always larger than in the one-dimensional case. It must also be noticed that keeping *r* constant, $t_{90\%}$ does not suffer violent variations with modifications in *P*. In particular, for low values of *r*, $t_{90\%}$ increases slightly with *P*. However, those enlargements seldom exceed 5% of the original value, and for example, for *r*=8 in two-dimensional movements, the variation of $t_{90\%}$ as *P* rises from 0.1 to 0.9 is below 4.25% ($t_{90\%}$ increases from 3.45 to 3.59 time units). With larger values of *r*, the variation of $t_{90\%}$ becomes negligible. For instance, for *r*=14000 in two-dimensional movements, the $t_{90\%}$ parameter remains constant at 3.6153 time units, for any value of *P* from 0.1 to 0.9.

Regarding $t_{98\%}$, this time interval is larger for the two-dimensional movement than for the one-dimensional case. Specifically, $t_{98\%}$ tends to vary slightly with *P*, although the main variation is presented with the values of *r*, increasing with it. In particular, the rise of this time interval with *r* is proportionally more important in the one-dimensional case than in the two-dimensional one. A brief summary of data concerning this parameter is presented in Table 7:

Table 7. Summary of measurements for the $t_{98\%}$ interval.

|   | One-dimensional movement | | | Two-dimensional movement | | |
|---|---|---|---|---|---|---|
| *P* | *r* =14 | *r* =1400 | *r* =14000 | *r* =14 | *r* =1400 | *r* =14000 |
| 0.5 | 3.86 | 4.344 | 4.39 | 5.63 | 5.64 | 5.64 |
| 0.1 | 4.08 | 4.18 | 4.33 | 5.596 | 5.646 | 5.647 |

In connection with the $t_{min}$ intervals, a reduction is measured for the two-dimensional movements in comparison with the one-dimensional case for the same values of *r* and *P*. For both types of movements, these time intervals diminish with rises in *r* or *P*. A selection of measurements is presented in Table 8:

Table 8. Summary of measurements for the $t_{min}$ interval.

|   | One-dimensional movement | | | Two-dimensional movement | | |
|---|---|---|---|---|---|---|
| *P* | *r* =14 | *r* =1400 | *r* =14000 | *r*=14 | *r* =1400 | *r* =14000 |
| 0.9 | 1.05 | 0.19 | 0.09 | 0.2 | 0.022 | 0.007 |
| 0.1 | >10 | 0.96 | 0.4 | 0.7 | 0.07 | 0.02 |

The same behavior is observed for values of $r \in$ [1.5, 13.5], where $t_{min}$ for the two-dimensional movement is around 1 time unit for $P$=0.1, while for $P$=0.9, it goes down to around 0.5 time units.

Table 9 gathers a summary of the ratios of the minimum obtainable costs between the two-dimensional and the one-dimensional movements, for different values of $r$ and $P$:

Table 9. Ratio of minima of costs between two-dimensional and one-dimensional movements, for timer method.

| RATIOS | P=0.1 | P=0.3 | P=0.5 | P=0.7 | P=0.9 |
|---|---|---|---|---|---|
| $r$ =1.4 | 3.5 | 3.14 | 2.5 | 2.3 | 2 |
| $r$ =5 | 5.24 | 3.57 | 3 | 2.7 | 2.5 |
| $r$ =8 | 5.82 | 3.6 | 3.14 | 2.8 | 2.6 |
| $r$ =14 | 6.28 | 4 | 3.3 | 3 | 2.8 |
| $r$ =1400 | 7.28 | 7.25 | 6.6 | 6 | 5.5 |
| $r$ =14000 | 12.4 | 10.06 | 9.08 | 8.51 | 8.08 |

From Table 9, it can be concluded that within the range of values of $r$ analyzed, the Location Management costs for the two-dimensional movements are always larger than for the one-dimensional case. It can also be noticed that the ratio for the minima between the two-dimensional and the one-dimensional case, diminishes as $P$ becomes larger, and for the same value of $P$, the ratio increases with $r$. An explanation for these trends is given next:

Assuming the amount of resources required to administer each Location Area as a constant independent of the movement dimension [1] (regardless of the exponential increase of the number of cells in the Location Area with the movement dimension), the higher the movement dimension, the lower the proportional amount of resources that would be required for the paging process. Figure 8 helps to illustrate this statement:

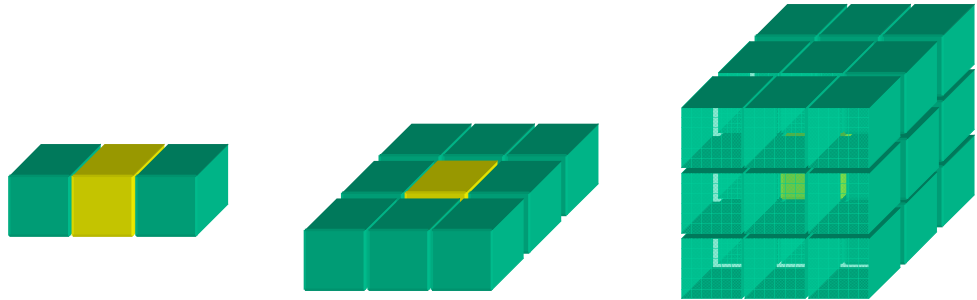

Figure 8. Variation of the mobile's specifiability with the movement dimension.

Mathematically, if $R$ is the total amount of resources required to administer a Location Area, the proportion of them theoretically used for a paging process by means of the timer-based method is higher the lower the movement dimension:

$$R_{used}\big|_{Paging} = \left(\frac{1}{3^n}\right) \cdot R \qquad (12)$$

Where n represents the movement dimension.

International Journal of Computer Networks & Communications (IJCNC) Vol.3, No.4, July 2011

Consequently, the paging costs term in the overall Location Management costs is proportionally more important for the one-dimensional movement than for the two-dimensional case. Therefore, increases in $P$ (cost of paging for each Location Area) bring proportionally larger rises in the average Location Management costs for the one-dimensional movements than for the two-dimensional ones. Hence, the ratio of minima between the two and one-dimensional movements diminishes with increases in $P$, as proven from the measurements carried out.

In opposition to the paging costs, the location update term in the overall Location Management costs is proportionally more important for the two-dimensional movement than for the one-dimensional case. Consequently, increases in the number of updates as a result of the rise in the value of $r$ make the Location Management costs grow proportionally more for the two-dimensional processes than for the one-dimensional ones. Therefore, the ratio of minima between the two and one-dimensional movements rises with $r$, as shown from the measurements carried out.

A change in the evolution of the ratio of minima of costs with $P$ beyond a certain threshold of $r$, would mean that the ratio would increase with $P$. However, checking the transition from $P = 0.1$ to $P = 0.3$ for a wide range of values of $r$, no variation is noticed in the decreasing behavior of the ratios with rises in $P$, as shown in Table 10:

Table 10. Ratio between minima of costs for two-dimensional and one-dimensional movements, for timer method, for $P=0.1$ and $P=0.3$, with $r > 500$.

| r | P=0.1 | | | P=0.3 | | |
| --- | --- | --- | --- | --- | --- | --- |
| | 2-D | 1-D | Ratio | 2-D | 1-D | Ratio |
| 500 | 17 | 2.04 | 8.333333 | 30 | 4.89 | 6.134969 |
| 800 | 21.7 | 2.5 | 8.68 | 38 | 5.83 | 6.51801 |
| 1000 | 24.5 | 2.7 | 9.074074 | 43 | 6.33 | 6.793049 |
| 1330 | 25.6 | 2.82 | 9.078014 | 45.1 | 6.56 | 6.875 |
| 1350 | 26.8 | 2.92 | 9.178082 | 47 | 6.77 | 6.942393 |
| 1380 | 27.8 | 3.02 | 9.205298 | 48.8 | 6.97 | 7.001435 |
| 2000 | 35 | 3.6 | 9.722222 | 61 | 8.25 | 7.393939 |
| 3000 | 43 | 4.2 | 10.2381 | 75 | 9.5 | 7.894737 |
| 4000 | 49 | 4.7 | 10.42553 | 87 | 10.5 | 8.285714 |
| 10000 | 79 | 6.5 | 12.15385 | 142 | 14.5 | 9.793103 |
| 12000 | 87 | 7.15 | 12.16783 | 155 | 15.4 | 10.06494 |
| 14000 | 93 | 7.6 | 12.23684 | 165 | 16.2 | 10.18519 |
| 30000 | 144 | 9.8 | 14.69388 | 235 | 21 | 11.19048 |
| 40000 | 161 | 10.9 | 14.77064 | 275 | 23.2 | 11.85345 |
| 50000 | 175 | 11.8 | 14.83051 | 305 | 25 | 12.2 |
| 100000 | 250 | 15 | 16.66667 | 450 | 32 | 14.0625 |
| 400000 | 500 | 24 | 20.83333 | 870 | 52 | 16.73077 |
| 500000 | 550 | 26 | 21.15385 | 960 | 55 | 17.45455 |
| 1000000 | 750 | 34 | 22.05882 | 1400 | 70 | 20 |
| 5000000 | 1800 | 57 | 31.57895 | 3100 | 120 | 25.83333 |
| 10000000 | 2500 | 72 | 34.72222 | 4300 | 151 | 28.47682 |

Nevertheless, for values of $r$ below 1, working with realistic time intervals of 10 time units, calculating the ratios for the minima in those time intervals, that ratio would increase with $P$, as observed in Table 11:

Table 11. Ratio between minima of costs for two-dimensional and one-dimensional movements, for timer method, for $P$=0.1 and $P$=0.3, with $r < 1$.

|  | P=0.1 |  |  | P=0.3 |  |  |
| --- | --- | --- | --- | --- | --- | --- |
| $r$ | 2-D | 1-D | Ratio | 2-D | 1-D | Ratio |
| 0.0014 | 0.0005 | 0.004 | 0.125 | 0.0015 | 0.011 | 0.136 |
| 0.014 | 0.0045 | 0.0125 | 0.36 | 0.013 | 0.035 | 0.37 |
| 0.14 | 0.044 | 0.038 | 1.157 | 0.13 | 0.11 | 1.18 |

However, the previous results must not lead us to the wrong conclusion, as it must be taken into account that the oscillation in the costs obtainable through the timer-based method tends to disappear for values of $r$ approximately below 1 for both movement dimensions, so that the costs would present an exponentially decreasing evolution at all times, in such a way that theoretically, the minimum of the costs would be reached for values of the time approaching $\infty$, instead of 10 time units. Consequently, making use of the real stable values calculated for the costs, the ratio for the minima between the two-dimensional and the one-dimensional movements would be $\pi \cdot \sqrt{r}$, independent of the value of $P$. Therefore, no increase with $P$ takes place in the evolution of the ratio of minima for any value of $r$.

## 6. CONCLUSIONS

We have analyzed the way in which the Location Management costs provided by the timer-based algorithm rise with the values of $r$ and $P$. In opposition to the timeout interval that minimizes the costs, $t_{90\%}$ grows with $r$ and $P$, and for typical values of these parameters, it increases with the movement dimension. These opposite behaviors of $t_{min}$ and $t_{90\%}$ can be helpful to adapt to different circumstances at the time of combining particular location update strategies with the timer-based method. The idea would be to apply a periodic update with a time interval which would interfere minimally with the other strategy, but still bringing lower costs than a "no update" policy. Depending on the specifications, a timeout interval longer than $t_{min}$ could be needed, for example $t_{90\%}$. Moreover, within the CMRs analyzed, the behavior of $t_{90\%}$ with modifications of $P$ proves to be very stable (measured variations under 5%).

For common values of $P$ below 0.3, taking CMRs lower than the 0.7 figure we have considered as threshold for a proper application of the timer-based method, the observed valley marking clearly the minimum in the Location Management costs loses importance for CMRs within [0.001, 0.7] for one-dimensional movements and within [0.2, 0.7] for the two-dimensional ones.

**Authors**

**E. Martin** received his PhD from England and is currently carrying out research in the University of California at Berkeley within the fields of wireless communications and signal processing.

**R. Bajcsy** received a PhD in EE from Czechoslovakia and another PhD in CS from Stanford (California). She is currently a Professor at the University of California, Berkeley. Amongst her interests are: body area sensor networks and wireless communications.